\documentclass[superscriptaddress,aps,prl,twocolumn,showpacs,nofootinbib, longbibliography, notitlepage,floatfix]{revtex4-2}
\usepackage{etex}
\usepackage{amsmath,amssymb,amsthm}
\usepackage[colorlinks=true,citecolor=blue,urlcolor=blue]{hyperref}
\usepackage[pdftex]{graphicx}
\usepackage{times,txfonts}
\usepackage{braket}
\usepackage{color}
\usepackage{natbib}
\usepackage{amsmath,blkarray}
\usepackage{mathtools}
\usepackage{ulem}
\usepackage{latexsym}
\usepackage{tabularx, booktabs}
\usepackage{graphics,epstopdf}
\usepackage{graphicx}
\usepackage{float}
\usepackage{amsfonts}
\usepackage{tikz}
\usetikzlibrary{quantikz}
\usepackage{color,soul}

\newcommand{\be}{\begin{equation}}
\newcommand{\ee}{\end{equation}}
\newcommand{\ba}{\begin{eqnarray}}
\newcommand{\ea}{\end{eqnarray}}

\usepackage{multirow}
\usepackage{appendix}
\usepackage{url}

\begin{document}
\title{Digitized-counterdiabatic quantum factorization} 

\author{Narendra N. Hegade}
\email{narendra.hegade@kipu-quantum.com}
\affiliation{Kipu Quantum, Greifswalderstrasse 226, 10405 Berlin, Germany}
\affiliation{International Center of Quantum Artificial Intelligence for Science and Technology~(QuArtist) \\ and Physics Department, Shanghai University, 200444 Shanghai, China}

\author{Enrique Solano}
\email{enrique.solano@kipu-quantum.com}
\affiliation{Kipu Quantum, Greifswalderstrasse 226, 10405 Berlin, Germany}

\date{\today}

\begin{abstract}
We factorize a 48-bit integer using 10 trapped-ion qubits on a Quantinuum's quantum computer. This result outperforms the recent achievement by B. Yan et al., arXiv:2212.12372 (2022), increasing the success probability by a factor of 6 with a non-hybrid digitized-counterdiabatic quantum factorization (DCQF) algorithm. We expect better results with hybrid DCQF methods on our path to factoring RSA-64, RSA-128, and RSA-2048 in this NISQ era, where the latter case may need digital-analog quantum computing (DAQC) encoding.
\end{abstract}
\maketitle
{\it Introduction.---} 
The recent proposal of Bao Yan et al. \cite{BaoYan}, inspired by the classical Schnorr's algorithm, shows that one could encode the integer factorization problem on a quantum computer with $\mathcal{O}(\log N / \log \log N)$ qubits, i.e., sublinear in the bit length of an integer N. In this sense, it is better than Shor's factorization algorithm \cite{shor1994algorithms} with respect to the number of qubits. However, the time complexity of this hybrid classical-quantum algorithm is unknown and hard to estimate. The authors combine Babai's algorithm with the quantum approximate optimization algorithm (QAOA) to solve the closest vector problem on a lattice. The resulting problem reduces to an optimization problem whose solution is encoded in the ground state of an Ising spin-glass Hamiltonian. Even though the authors experimentally factorize a 48-bit number on a superconducting quantum computer with a large enough success probability, its scalability for larger tasks remains unknown.   
In this report, we take up the classical preprocessing part of their work and enhance the quantum part of the algorithm. We propose a non-hybrid approach, called digitized-counterdiabatic quantum factorization (DCQF), to tackle the same problem outperforming QAOA techniques. In this sense, DCQF may allow us to factorize larger numbers, possibly up to RSA-64 and RSA-128, with current noisy intermediate-scale quantum (NISQ) computers. This report explores the possibility of factoring larger numbers with compressed algorithms in given quantum computers, rather than proving any scalability of computational resources. On the other hand, further use of analog~\cite{ana1, MISQuera} or digital-analog encoding schemes \cite{daqc1, daqc2, AMichel} may pave the way towards factoring RSA-2048 in the NISQ era, without the long wait for fault-tolerant quantum computers.

In Ref.~\cite{BaoYan}, the quantum part of their factorization algorithm consists in finding the ground state on an Ising spin-glass Hamiltonian with all-to-all connectivity. The general form of such Hamiltonian is given by
\begin{equation} \label{eq1}
H_{Ising} = \sum_{i<j} J_{i j} \sigma_i^z \sigma_j^z+\sum_i h_i \sigma_i^z.
\end{equation}
Here, $\sigma_i^z$ is the Pauli-$z$ matrix, $J_{i j}$, and $h_i$ are the interactions between the spins and local field acting on a site $i$, respectively. 
In the worst-case scenario, finding the ground state of an Ising spin-glass problem is known to be NP-hard. Along these lines, even with quantum computers, it is unlikely to solve this problem in polynomial time, though one could expect a polynomial quantum speed-up. There are various approaches to tackle this problem on a quantum computer, using adiabatic quantum computation (AQC), quantum annealing (QA) \cite{albash2018adiabatic}, QAOA \cite{qaoa}, among others. Despite the vast interest in QA and QAOA for solving combinatorial optimization problems, we still need to learn about their quantum speed-up for large-scale problems of industrial relevance. Here, we will consider digitized-counterdiabatic quantum computing (DCQC) applied to the factorization problem, which is known to overcome some of the challenges faced by AQC and to outperform QAOA~\cite{hegade2021shortcuts, cdqaoa2, proteinfolding}.

Counterdiabatic (CD) protocols are known to speed up the adiabatic evolution by suppressing the non-adiabatic transitions. The recent developments in this field have opened the possibility of applying these techniques to AQC \cite{sels2017minimizing, claeys2019floquet}. Even with approximate counterdiabatic terms, a drastic enhancement can be obtained for most problems \cite{hegade2021shortcuts, cdqaoa2, proteinfolding, sels2017minimizing, claeys2019floquet, hegade2022DCQO, poly2, CDhubbard}. However, the experimental implementation of the CD protocols on analog quantum computers is a challenging task. Especially, while solving classical optimization problems, the CD terms are shown to be non-stoquastic, and the current quantum annealers do not have the capability to consider such problems. In order to overcome these difficulties, DCQC was proposed and experimentally tested. Recently, it was numerically proved that even the simplest approximate CD protocols could offer polynomial scaling enhancement in the ground state success probability, as compared with the finite time adiabatic quantum optimization~\cite{hegade2022DCQO, poly2}.

{\it DCQF algorithm.---} In order to find the ground state of Hamiltonian in Eq.~\eqref{eq1}, we start with an adiabatic Hamiltonian defined as
\begin{equation} \label{eq2}
H_{ad}(\lambda)=[1-\lambda(t)] H_i+\lambda(t) H_{Ising},
\end{equation}
where $\lambda(t)$ is a scheduling function which defines the path between $H_i$ and $H_{Ising}$. We choose $\lambda(t) = \sin ^2\left[\frac{\pi}{2} \sin ^2\left(\frac{\pi t}{2 T}\right)\right]$ such that its first and second derivatives vanish at the initial and final time. This is an optional boundary condition for the CD protocol. The initial Hamiltonian is chosen as $H_i = -\sum_i \sigma^x_i$ such that its ground state $\ket{+}^{\otimes n}$ can be easily prepared. In order to speed-up the adiabatic evolution, we introduce an approximate CD term as $H(\lambda) = H_{ad}(\lambda) + \dot{\lambda} A_\lambda^{(l)}$. Here, $A_\lambda^{(l)}$ is the approximate adiabatic gauge potential (AGP) obtain from nested commutator expansion \cite{claeys2019floquet} given by
\begin{equation}
A_\lambda^{(l)}=i \sum_{k=1}^l \alpha_k(\lambda) \underbrace{\left[H_{a d},\left[H_{a d}, \ldots\left[H_{a d}\right.\right.\right.}_{2 k-1}, \partial_\lambda H_{a d}]]]. 
\end{equation}
Here, the CD coefficients $\alpha_k(\lambda)$ can be obtained by minimizing the action $S=\operatorname{Tr}\left[G_\lambda^2\right]$, where $G_\lambda=\partial_\lambda H_{a d}+i\left[A_\lambda^{(l)}, H_{a d}\right]$ is a Hermitian operator. Recently, there have been some alternative approaches to deterministically obtain $\alpha_k(\lambda)$ by solving a set of linear equations \cite{CDhubbard}.    
As we consider higher order expansion terms $l$, we get better approximations of the exact AGP. However, for large $l$, we will get long-range multi-qubit interactions, resulting in large circuit depth. Therefore, in this work, we set $l=1$.

To show the performance of DCQF algorithm, we consider the same Hamiltonians as in Ref.~\cite{BaoYan} corresponding to factoring 26-bit and 48-bit numbers with 5 and 10 qubits, respectively. We start with the 26-bit number 48567227. After the classical preprocessing, the problem reduces to find the ground state of a 5-qubit Hamiltonian given by 
\begin{equation}
\begin{aligned} \label{eq3}
H_{5q} &=  781 \mathbb{I}-142 \sigma_z^1-64 \sigma_z^2-81 \sigma_z^3-213 \sigma_z^4-4.5 \sigma_z^5 \\ & -13.5 \sigma_z^1 \sigma_z^2+3.5 \sigma_z^1 \sigma_z^3+18 \sigma_z^1 \sigma_z^4+17.5 \sigma_z^1 \sigma_z^5 \\ & -29 \sigma_z^2 \sigma_z^3+19.5 \sigma_z^2 \sigma_z^4-34 \sigma_z^2 \sigma_z^5-31.5 \sigma_z^3 \sigma_z^4\\ &  -2.5 \sigma_z^3 \sigma_z^5+4.5 \sigma_z^4 \sigma_z^5 \; .
\end{aligned}
\end{equation}
\begin{figure}
    \centering
    \includegraphics[width=\linewidth]{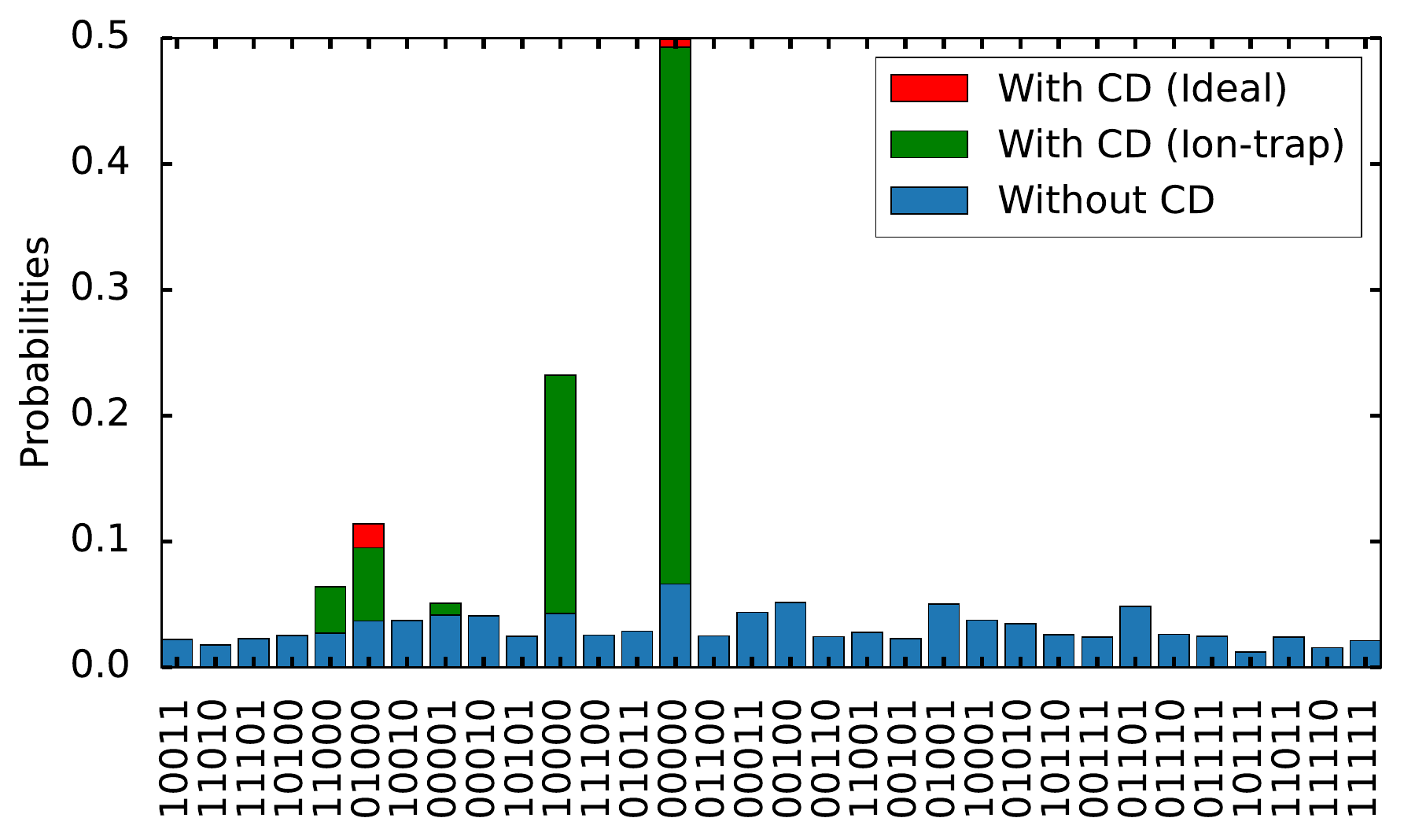}
    \caption{Probability distribution obtained from DCQF evolution for factoring the number 48567227 (26-bit) with 5 qubits. We considered total evolution time $T=0.4$ and $dt=0.1$ with and without including the CD term. The probability of obtaining the ground state using the DCQF protocol has a larger overlap with the ground state ($\approx 50\%$) for both ideal and experimental results in Quantinuum's trapped-ion quantum processor, where $1000$ shots were made.}
    \label{fig1}
\end{figure}
To solve this problem, we consider a simple local AGP $\tilde{A}_\lambda=\sum_i^{n=5} \beta_i(t) \sigma_i^y$, where the CD coefficient is calculated as $\beta_i(t)=h_i / 2\left[(\lambda-1)^2+\lambda^2\left(h_i^2+\sum_{j\neq i}^n J_{i j}^2\right)\right]$. The total Hamiltonian including the CD term is $H(\lambda) = [1-\lambda(t)] H_i+\lambda(t) H_{5q} + \dot{\lambda} \tilde{A}_\lambda$. For a fast evolution, the CD term plays a dominant role so that $H(\lambda) \approx \dot{\lambda} \tilde{A}_\lambda$. Interestingly, the resulting Hamiltonian contains only local one-body terms. To verify the performance of this local CD term, we digitized the time evolution using the first-order Trotter-Suzuki formula. We consider the total evolution time $T=0.4$ and time step $dt=0.1$, that is four Trotter steps. The resulting unitary operator is decomposed into a set of quantum gates. For experimental demonstration, we consider Quantinuum's 20-qubit trapped-ion processor and make use of their native gates. The final state of the system at the end of the evolution is extracted by measuring in the computational basis. In Fig.~\ref{fig1}, the probability distribution obtained from both the ideal simulator and the experimental results are plotted. We notice that, even with the short evolution time, the CD protocol is able to find the ground state $\ket{00000}$ corresponding to the Hamiltonian \eqref{eq3} with $\approx 50\%$ success probability from the ideal simulator, and $49.3\%$ from the experiment.  We remark that, for Hamiltonians with $|h_i| >> |J_{ij}|$, even the local CD terms can provide a drastic enhancement in the ground state success probability. However, even for such simple instances, the result obtained in reference \cite{BaoYan} using QAOA with $p=3$ is not able to find the correct solution with $>30\%$ success probability.
\begin{figure*}
    \centering
    \includegraphics[width=\linewidth]{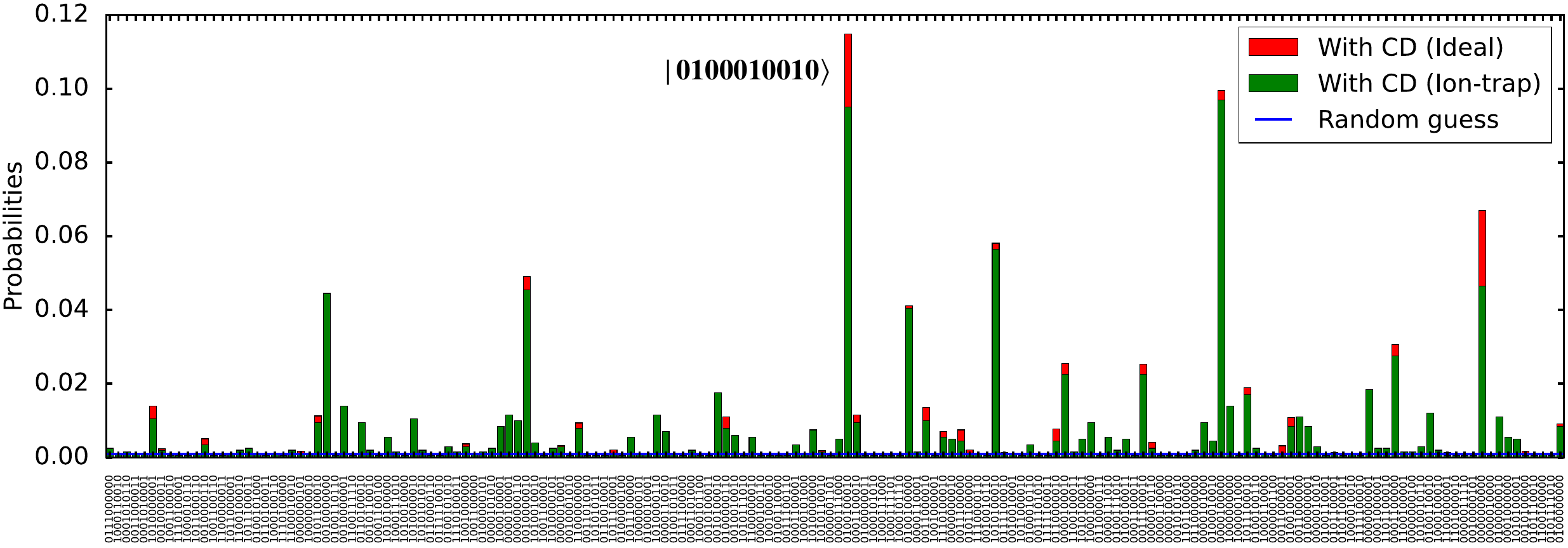}
    \caption{Final probability distribution obtained with DCQF for factoring the 48-bit integer 261980999226229 with 10 qubits on Quantinuum's trapped-ion quantum computer. The solution of the problem is encoded in the ground state of an all-to-all connected Ising spin-glass Hamiltonian. An approximate 2-local CD term is added to speed up the evolution. Both the theoretical and the experimental results show that the ground state $\ket{0100010010}$ is obtained with a higher success probability. We considered 2000 shots for this experiment.}
    \label{fig2}
\end{figure*}
As a second example, we consider factoring the 48-bit integer 261980999226229 using 10 qubits. The Hamiltonian encoding the solution of the problem is obtained from Ref.~\cite{BaoYan}. Unlike the previous case, here, the local fields $h_i$ and the interaction terms $J_{ij}$ have comparable magnitudes. So, we consider the first-order nested commutator in Eq.~\eqref{eq2} to obtain the approximate CD terms. For a fair comparison with QAOA, we further truncate the number of 2-local terms in AGP to match the circuit depth of QAOA for each layer $p$. The resulting CD Hamiltonian is given by  
$H_{c d}^{(1)}(\lambda)=-4 \dot{\lambda} \alpha_1(\lambda)\left[\sum_i h_i \sigma_i^y+\sum_{i<j} J_{i j}\sigma_i^z \sigma_j^y\right]$. The exact solution for the first-order CD coefficient $\alpha_1(\lambda)$ is given in Ref.~\cite{hegade2022DCQO}. As before, we consider fast evolution with $H(\lambda) \approx H_{c d}^{(1)}(\lambda)$. The total evolution time is $T=0.4$ and time step $dt=0.1$. Even though there are 4 Trotter steps, the CD term vanishes at the $t=T$ due to the boundary condition $\dot{\lambda}(t=0) = \dot{\lambda}(t=T)=0$. Also, the magnitude of the CD terms reduces rapidly after one Trotter step. So, effectively the contribution to the time evolution comes from the first Trotter step, and one could discard the CD terms in the second and third Trotter steps with very small gate angles. The resulting time evolution operator contains in total 45 two-body interaction terms. For the experimental implementation, we decompose the interaction terms using the native $ZZ (\theta)$ gates. In Fig~\ref{fig2}, we show the final probability distribution obtained from the CD protocol using an ideal simulator as well as the experimental result from the trapped-ion system. We can see that the ground state $\ket{0100010010}$ is obtained with an $11.5\%$ probability on an ideal simulator whereas the experimental probability is 9.7\%. For the same problem, p=1 QAOA is able to give a theoretical ground state success probability of 2\%. Even with p=3, QAOA is not able to go beyond 4\% success probability. As an additional note, we observed an increase in minimum energy gap $\Delta_{min}$, i.e., the minimum energy gap between the ground state and the first excited state during the evolution by including the CD terms. A detailed analysis on such observation can be found in our previous work \cite{hegade2022DCQO}.

{\it Conclusion.---} 
We showed how to factor a 48-bit integer on Quantinuum's trapped-ion quantum processor with the DCQF algorithm. It does not require any classical optimization routines and, at the same time, outperforms QAOA in finding the ground state. This means we could still explore the application of hybrid DCQF methods on current hardware, enhancing the presented performance. We estimate that, with a hybrid-DCQF algorithm, one could factor RSA-64 on current NISQ computers with around 20 qubits. Also, with optimized hardware adaptations, we consider that one could factor RSA-128 with 37 qubits in the near future. Going beyond that with current digital quantum computers, for given gate fidelities and coherence times, looks like a challenging task. However, alternative approaches like digital-analog quantum computing, involving multiqubit analog blocks in combination with local digital steps, might tackle RSA-2048, bringing quantum advantage to the NISQ era~\cite{daqc1, daqc2, AMichel}. Moreover, one could also consider this problem on programmable analog quantum simulators, like neutral atom devices \cite{pasqal,MISQuera}, with Floquet engineering techniques. In this manner, one may effectively realize the counterdiabatic protocol for factoring RSA-2048 without sophisticated error mitigation. We believe this report suggests a path towards quantum advantage for industry problems with current quantum computers.


\begin{thebibliography}{32}
\bibitem{BaoYan}B. Yan {\it et al.}, \href{https://arxiv.org/abs/2212.12372}{arXiv:2212.12372 (2022).}

\bibitem{shor1994algorithms}P.~W.~Shor, in \href{https://ieeexplore.ieee.org/document/365700}{Proceedings of the 35th Annual Symposium on Foundations of Computer Science} (IEEE Computer Society Press, Los Alamitos, 1994), pp. 124.

\bibitem{ana1} A.~J.~Daley {\it et al.}, \href{https://www.nature.com/articles/s41586-022-04940-6}{Nature \textbf{607}, 667–676 (2022).}

\bibitem{MISQuera} S.~Ebadi {\it et al.}, \href{https://www.science.org/doi/10.1126/science.abo6587}{Science \textbf{376}, 6587 (2022).}

\bibitem{daqc1} A.~Parra-Rodriguez {\it et al.}, \href{https://journals.aps.org/pra/abstract/10.1103/PhysRevA.101.022305}{Phy. Rev. A \textbf{101}, 022305 (2020).}

\bibitem{daqc2} D.~Headley {\it et al.}, \href{https://journals.aps.org/pra/abstract/10.1103/PhysRevA.106.042446}{Phy. Rev. A \textbf{106}, 042446 (2022).}
\bibitem{AMichel}A.~Michel {\it et al.}, \href{https://arxiv.org/abs/2301.06453}{arXiv:2301.06453 (2023).}

\bibitem{albash2018adiabatic}T.~Albash and D.~A.~Lidar,  \href{https://journals.aps.org/rmp/abstract/10.1103/RevModPhys.90.015002}{Rev. Mod. Phys. \textbf{90}, 015002 (2018).}

\bibitem{qaoa}E.~Farhi, J.~Goldstone, and S.~Gutmann, \href{https://arxiv.org/abs/1411.4028}{arXiv:1411.4028 (2014).}

\bibitem{hegade2021shortcuts}N.~N.~Hegade {\it et al.}, \href{https://journals.aps.org/prapplied/abstract/10.1103/PhysRevApplied.15.024038}{Phys. Rev. Appl. \textbf{15}, 024038 (2021).} 

\bibitem{cdqaoa2}J.~Yao, L.~Lin, and M.~Bukov, \href{https://journals.aps.org/prx/abstract/10.1103/PhysRevX.11.031070}{Phys. Rev. X \textbf{11}, 031070 (2021).} 

\bibitem{proteinfolding}P.~Chandarana {\it et al.}, \href{https://arxiv.org/abs/2212.13511}{arXiv:2212.13511 (2022).}

\bibitem{CDhubbard}Q.~Xie, K.~Seki, and S.~Yunoki, \href{https://journals.aps.org/prb/abstract/10.1103/PhysRevB.106.155153}{Phys. Rev. B \textbf{106}, 155153 (2022).}

\bibitem{sels2017minimizing}D.~Sels and A.~Polkovnikov, \href{https://www.pnas.org/content/114/20/E3909}{PNAS \textbf{ 114}, E3909 (2017).}

\bibitem{claeys2019floquet}P.~W.~Claeys, M.~Pandey, D.~Sels, and A.~Polkovnikov, \href{https://journals.aps.org/prl/abstract/10.1103/PhysRevLett.123.090602}{Phys. Rev. Lett. \textbf{123}, 090602 (2019).}

\bibitem{hegade2022DCQO} N.~N.~Hegade, X.~Chen, and E.~Solano, \href{https://journals.aps.org/prresearch/abstract/10.1103/PhysRevResearch.4.L042030}{Phys. Rev. Res. \textbf{4}, L042030 (2022).}

\bibitem{poly2} A.~Hartmann, G.~B.~Mbeng, and W.~Lechner, \href{https://journals.aps.org/pra/abstract/10.1103/PhysRevA.105.022614}{Phys. Rev. A \textbf{105}, 022614 (2022).}

\bibitem{pasqal} L.~Henriet {\it et al.}, \href{https://quantum-journal.org/papers/q-2020-09-21-327/}{Quantum \textbf{4}, 327 (2020).}

\end{thebibliography}
\end{document}